\documentstyle[pra,aps,preprint]{revtex}
\begin{document}
\title{Confinement and Bose Condensation in
Gauge Theory of High-Tc Superconductors}
\author{Naoto Nagaosa}
\address{Department of Applied Physics, University of Tokyo,
Bunkyo-ku, Tokyo 113, Japan}
\author{Patrick A. Lee}
\address{Department of Physics, Massachusetts Institute of Technology,
Cambridge, MA02139, U.S.A.}
\date{\today}
\maketitle
\begin{abstract}
The issue of confinement and bose condensation
is studied for gauge models of high-Tc superconductors.
First the Abelian-Higgs model in (2+1)D, i.e.,
XY-model coupled to lattice gauge field $a_{\mu}$ with coupling $g$,
is studied taking into account both the
instantons and vortices.
This model corresponds to  integer filling of the bosons,
and can be mapped to a
dual superconductor. Our main resut is that the instantons introduce a term
which couples linearly to the dual superconductor order parameter, and tend to
pin its phase. As a result
the vortex condensation always occurs due to the instantons, and
the Meissner effect for the gauge field $a_\mu$ is absent,
although $a_{\mu}$ is massive. This state is essentially the same as the
confining phase of the pure gauge model.
Away from  integer filling, a ``magnetic field'' $\mu$
(the chemical potential of the bosons) is applied to this dual superconductor.
Then the Higgs phase revives in the case of weak $g$ and large $x$, where
vortices do not condense in spite of the instantons.
In the opposite case, i.e., strong  $g$ and small $x$,
phase separation occurs, forming either microscopic patches or macroscopic
stripe domains of the Mott insulating state.
\end{abstract}
\narrowtext
\section{Introduction}
It has been established that the strong Coulomb repulsion between
electrons is the key issue in the physics of high-Tc cuprates.
Anderson proposed that this strong correlation gives rise to
resonating valence bond (RVB) state, where the spin and charge are carried
by spinons and holons, respectively \cite{and1}.
This phenomenon is called spin-charge separation, and
has been subject to intensive studies.
One way of formulating the strong electron repulsion
is to exclude the double occupancy of the electrons on each site,
and study the effective Hamiltonian within that restricted Hilbert space.
Slave boson method is a useful tool to implement this constraint
and fits the idea of spin-charge separation,
where two species of particles, i.e., spinons
(fermions $f^\dagger_{i \sigma}$, $f_{i \sigma}$)  and
holons (bosons $b_i^\dagger$, $b_i$), are introduced to represent electron
operator $C^\dagger_{i \sigma}$, $C_{i \sigma}$ as
\cite{bas1,kot,suzu}
\begin{eqnarray}
C^\dagger_{i \sigma} &=& f^\dagger_{i \sigma} b_i,
\nonumber \\
C_{i \sigma} &=& f_{i \sigma} b^\dagger_i.
\end{eqnarray}
Here an electron is represented as the composite particle of
spinon and holon. At the mean field level, these two species of particles are
supposed to be independent of each other.
The mean field phase diagram is determined by two phase transition lines,
i.e, the spinon pairing transition and holon condensation
characterized by the order parameters $\Delta = < f_{i \sigma} f_{j \sigma}>$
and $B = <b_i>$, respectively. \cite{bas1,kot,suzu}
In strange metal state both $\Delta$ and $B$ are zero, while
only $\Delta$ is nonzero in the underdoped ``spin gap state''  and
only $B$ is nonzero in the overdoped ``Fermi liquid state''.
The superconductivity is realized only when both $\Delta$ and $B$
are nonzero, and the onset of the superconductivity is
identified as the holon condensatoin in the underdoped region.
However this simple-minded picture of spin-charge separation needs to be
critically studied because the constraint is replaced by the average one
in the mean field theory and the more appropriate treatment of this
constraint might change the whole picture.
This constraint can be taken care of by gauge field which
corresponds to the fluctuation of the phase of RVB order parameter
and Lagrange multiplier to impose the constraint \cite{bas,ioffe,nag1}.
Therefore the effective model is that of the spinons and holons
coupled to the gauge field.
It should be noted here that the gauge field represent the constraint and
does not have its own dynamics. The effective action and the
inverse of the coupling constant $1/g$
for the gauge field is generated only after integrating over spinons
and holons.
Therefore it is a highly nontrivial and crucial issue if the weak coupling
perturbative analysis with respect to $g$ makes sense or not.

This issue is closely related to the confinement/deconfinement of the gauge
field.\cite{bas,laughlin1,laughlin2} The original model is defined on a
lattice and the gauge field
is compact, and it is well-known that the gauge field is confining
in the strong coupling limit, i.e., large $g$. \cite{kogut}
A simplified and rough picture of the confinement follows.
On the lattice one can define the gauge flux $b(p)$ penetrating each plaquette
$p$, and the action is periodic with respect to $b(p)$ with
period $2 \pi$.  The simplest potential energy for $b(p)$ is
$ - g^{-1} \cos b(p)$, and the kinetic energy of $b(p)$ is given by
${ g \over 2} e^2$ where $e$ is the electric field canonical conjugate to
$b(p)$. 
A phase transition is possible between  two states.
One is the extended ``Bloch wave state'' of $b(p)$ where the tunneling events
between different minima of the poriodic potential are driven by the large
kinetic energy, i.e., large $g$. The conjugate field $e$ is localized
on the other hand, and the string of the electric field is formed
when positive and negative gauge charges are inserted with a separtion $R$.
This costs an energy proportional to $R$ because of the finite string tension
of the electric field. This phenomenon is called confinement.
For small $g$, on the other hand, the periodic potential is
large and $b(p)$ is confined within one minima.
One can replace the poriodic potential by the quadratic one
${ 1 \over {2g}} b^2$, which corresponds to the usual Maxwell Lagrangian.
The Coulomb law is reproduced in this case, and the periodicity is
irrelevant.  The gauge field is deconfining in this case.
In (2+1)D these tunneling events are represented by the point singularities
of the gauge field configuration
called magnetic monopoles or instantons.\cite{pol} In the pure gauge model,
the interaction between the magnetic minopoles is $1/r$, i.e.,
the Coulomb gas.
Because the Coulomb gas in (2+1)D is in the screening phase,
the gauge field is always confining due to these monoploes/instantons.

When coupled to the holons and spinons, however, the gauge field
becomes dissipative when these particles are integrated over,
and the deconfining phase becomes possible in the strange metal
normal state.\cite{nag2} However , in the presence of gapless fermion or boson
excitations, the integration over them is in general not justifiable. The
fermion part is
better controlled because of the presence of a large energy scale, the Fermi
energy, and
techniques such as large N can be used to control the expansion. The boson
part is much
more problematic because bosons tend to condense in the bottom of their
band and we are
faced with a strong coupling problem of bosons and gauge fields.

It has been argued that the strong inelastic scattering due to
gauge fluctuation suppresses the coherency and hence the ordering
temperature, but the effects of the quantum fluctuation in the strong
coupling limit is a
difficult problem which remains unsolved. In this paper we hope that a
duality mapping of
a simplified version of the boson gauge field problem can shed some light
on the issue.
Summarizing the above, there are two crucial and related issues
in gauge models of high-Tc superconductors, i.e.,
the Bose condensation and the confinement/deconfinement of the gauge field.

In order to clarify these two pictures, we study in this paper a simplified
model, i.e.,
a Higgs model coupled to U(1) gauge field defined on a
(2+1)D lattice,
which is an important model of broad interests both in condensed matter physics
and high energy physics.\cite{frad,savit}
The action is given as
\begin{equation}
S = - \kappa \sum_{\rm link } \cos [ \Delta_\mu \theta (i) - q a_{\mu}(i) ]
    - { 1 \over g } \sum_{\rm plaquette} \cos
     [ \Delta_\mu a_\nu (i) - \Delta_\nu a_{\mu}(i) ]
\end{equation}
where $i$ is the lattice point and $\mu,\nu$ specify the direction
in the (2+1)D lattice.
The difference operator $\Delta_\mu$ is defined as
$\Delta_\mu f(i) = f( i + \mu) - f(i)$.
Both $\theta(i)$ and $a_\mu(i)$ are compact
and defined in the interval $[0, 2\pi]$.
$q$ is the (integer) charge  of the Higgs bosons and
$q=e$ (fundamental) for the holons while $q=2e$ for the spinon pairing
in the U(1) gauge model of high-Tc cuprates \cite{nag1}.
Hereafter we take the unit $e=1$ except for eq.(27) in section IV.
The amplitude of the bose field has been fixed, but the vortex
excitations are allowed due to the lattice and the compactness.  Note that,
in contrast to the high
$T_c$ problem, the gauge field dynamics is Maxwellian in the continuum
limit. This model has been
studied extensively, and the essential features are as follows.

\begin{description}
\item [(1)] In the limit $g \rightarrow 0$, the gauge fields are frozen out
and we recover the XY-model.  For $d+1 \geq 2$, we expect a phase
transition at $\kappa =
\kappa_c$.  The ordered phase in $d+1 \geq 3$ is
characterized by an order parameter $<e^{i\theta}> \neq 0$.  For $g$
nonzero, however, this object is not
gauge invariant and can no longer serve as an order parameter.
\cite{eli}  Instead we
may fix the gauge to be, for
instance, the unitary gauge $\theta = 0$ and consider small gauge
fluctuations.  Then we have the Higgs
mechanism where a term $\rho_sa^2$ appears in the action
( $\rho_s$: the superfluidity density).
  The gauge flux correlation has the form
\begin{equation}
<b_\mu b_\nu > = \frac{k^2}{\rho_s} \left( \delta_{\mu\nu} - \frac{k_\mu
k_\nu}{k^2} \right) \,\,\,\, .
\end{equation}
\item[(2)]  The limit $\kappa = 0$ yields the  compact QED
model in $2+1$ dimensions, which is known to be
confining due to the appearance of instantons.\cite{pol}  The instantons
are singular configurations of
$\vec a$ which act as magnetic monopoles, i.e., sources and sinks of
magnetic fields.  Writing $\vec a
= \vec a_0 + \vec a_{\rm inst}$ as the sum of nonsingular and singular
configurations, we have
\begin{equation}
\vec b_{\rm inst} = \vec \nabla \times \vec a_{\rm inst}
\end{equation}
For a given instanton density $\rho_M$, we have the analog of Poisson's
equation,
\begin{equation}
\vec \nabla \cdot \vec b_{\rm inst} =
\vec \nabla \cdot \vec \nabla \times \vec
a_{\rm inst} = 4\pi \rho_M
\,\,\,\, .
\end{equation}
If the instantons form a gas, the density-density correlation function is
given by
\begin{equation}
<\rho_M(\vec k)\rho_M(-\vec k) > = \frac{M_0^2k^2}{M_0^2 + k^2}
\end{equation}
in analogy with the more familiar density-density correlation function of
a Coulomb gas, where $M_0$
plays the role of the inverse screening length.  Combining Eq. (4) and Eq.
(5) we find
\begin{equation}
<b_{\rm inst}^\mu b_{\rm inst}^\nu >
= \frac{k_\mu k_\nu}{k^2} \frac{M_0^2}{M_0^2 + k^2}
\,\,\,\, .
\end{equation}
When combined with the standard transverse correlation from the nonsingular
part $\vec a_0$, we have
\begin{eqnarray}
<b_\mu b_\nu > & = & \delta_{\mu \nu} - \frac{k_\mu k_\nu}{k^2} + 
\frac{k_\mu k_\nu}{k^2} \frac{M_0^2}{M_0^2 + k^2}
\nonumber
\\
         & = & \delta_{\mu \nu} - \frac{k_\mu k_\nu}{M_0^2 + k^2}
\end{eqnarray}
Equation (7) shows that the electromagnetic field acquires a mass due to
the gas of instantons.\cite{pol}
\end{description}

Thus we see that in both the $g \rightarrow 0$ limit and the $\kappa = 0$
limits, the gauge field is
massive.  This inspired Fradkin and Shenker\cite{frad} to consider whether
the two limits are smoothly
connected to each other.  Their conclusion is as follows:

\begin{description}
\item [(1)]  When the charge is fundamental ($q=1$), the strong
coupling expansion argument shows that the Higgs phase with
large $\kappa$ and small $g$ continues smoothly to the
confinement phase with large $g$ and small $\kappa$.
\item[(2)] When the charge is not fundamental, e.g., $q=2$, the
Higgs  and the confinement are different phases and can be distinguished
by the forces between the test charges $q_0 = \pm 1$.
\end{description}

In the bulk of this paper we will focus on the $q = 1$ case.
The phase diagram for (3+1)D and $q=1$ case has been determined at least
qualitatively, and consists of two phases, i.e., the Higgs-confinement
and Coulomb phases as shown in Fig. 1(a).
In the Coulomb phase, the gauge field is deconfining and massless
and the bose field remains disordered.
However the phase diagram for (2+1)D and $q=1$, which is the most relevant
case to high-Tc cuprates,  remains controversial \cite{savit,frad}.
One is that the XY-transition becomes first order once the
coupling $g$ to the gauge field is turned on, due to the Weinberg-Coleman
mechanism
\cite{hal}.
This first order transition line terminates at some critical point. This
resembles the vapor-liquid phase diagram.
The other one is that a finite region of Coulomb phase exists.\cite{savit}

Since instantons dominate the physics for $\kappa = 0$, the key question is
whether instantons play an
important role for general $\kappa$ and $g$.  Vortex lines in the bose
liquid carry unit flux quanua
and these can originate and terminate at instantons and anti-instantons.
This is illustrated in Fig.~2.
Einhorn and Savit discussed the free energy of the finite vortex segments,
and concluded that the XY-transition remains at least for weak $g$.
\cite{savit}  This leads to a picture of a
finite region of the Coulomb
phase mentioned earlier.  Apart from the phase diagram, another issue is
the behavior of the magnetic
field correlation function in the region commonly labelled as
confinement-Higgs after the work of Fradkin
and Shenker. \cite{frad}
The point is that even though the gauge field is massive, the
correlation function in the
Higgs phase and the confinement phase are very different, being given by
Eq.~(2) and Eq.~(7),
respectively.  The question is then whether the correlation function is
closer to Eq.~(2) or Eq.~(7) in
the confinement-Higgs region.  As far as we know, this question has not
been addressed.  In this paper we
include the effect of instantons for general $\kappa$ and $g$ but we come
to quite a different conclusion
than Einhorn and Savit.\cite{savit}
The phase diagram we propose is shown in
Fig.~1(b).  The line $g=0$ is an
isolated line with an isolated XY-transition.  We also calculate the
magnetic field correlation function
and show that it takes the confinement form of Eq.~(7).  Thus we conclude
that the Higgs-confinement
phase is better described as the confinement phase.  This is the main
result of this paper which we
discuss in Section II.

It is known that the model given by Eq.~(1) describes a Bose gas on a
lattice where the density is an
integer.  Thus the confinement phase may be understood as a Bose Mott
insulator.  In Section III we
extend the discussion to the incommensurate case, when the density is not
an integer.  We find two
possibilities.  The instantons may become irrelevant and the system becomes
a superfluid (Higgs phase),
or the system may break up into domains.  In Section IV we briefly
address the case $q = 2$ and
also the case of two kinds of bosons with opposite gauge charges, which
arises in the SU(2) formulation
of the $t$-$J$ model.

\section{The Abelian-Higgs Model}

We start with the continuum action for eq.(1).
\begin{equation}
S = \int d^3 x \biggl[ { 1 \over 2} \kappa ( \nabla \theta(x) - \vec a(x)
- \vec A(x) )^2
 + { 1 \over { 2g}} ( \nabla \times \vec a )^2 \biggr]
\end{equation}
Here we have introduced the external electromangetic field $A_\mu$,
which is put to be zero for the moment.
We allow the singular configurations of $\theta$ and $\vec a$
corresponding to the vortex and instanton, respectively.
As for the vortex, $\theta(x)$ is devided into the single-valued part
$\theta_0$ and the multi-valued vortex part $\theta_V$ as
$\theta = \theta_0 + \theta_V$, and
$\nabla \times \nabla \theta_V/ 2 \pi  = \vec j_V$ is the vortex current.
As for the instantons,
$\vec a = \vec a_0 + \vec a_{\rm inst}$ and
$\nabla \cdot \nabla \times a_{\rm inst} /4 \pi =\rho_M$ is the
instanton density as defined in Eqs. (4) and (5). We next derive a
continuum version
of the duality representation of this problem. The duality representation is
 a powerful tool which allows us to discuss the strong coupling limit 
( large $g$ ) and
 is particularly useful for the present problem.
Introducing the Stratonovich-Hubbard field $\vec J$ representing the
boson current, the action becomes
\begin{equation}
S = \int d^3 x \biggl[ { 1 \over 2 \kappa} \vec J^2 + i \vec J \cdot
 ( \nabla \theta(x) - \vec a(x) )
 + { 1 \over { 2g}} ( \nabla \times \vec a )^2 \biggr]
\end{equation}
and after integrating over $\theta_0$, we obtain $\nabla \cdot \vec J =0$
corresponding to the conservation of the boson current.
To enforce this conservation law, we introduce the vector potential $\vec
c$ to represent
$\vec J$ as $\vec J = \nabla \times \vec c$.
Then after partial integration the action becomes
\begin{equation}
S = \int d^3 x \biggl[ { 1 \over 2 \kappa} (\nabla \times \vec c)^2 +
i \vec c \cdot
 ( 2 \pi \vec j_V - \vec b(x) )
 + { 1 \over { 2g}} (\vec b)^2 \biggr]
\end{equation}
with $\vec b = \nabla \times \vec a$.
After integrating over $\vec b$, we obtain
\begin{equation}
S = \int d^3 x \biggl[ { 1 \over 2 \kappa} (\nabla \times \vec c)^2 +
{ g \over 2} ({\vec c})^2
+ i 2 \pi \vec c \cdot \vec j_V \biggr]
\end{equation}
This is essentially equivalent to the action obtained by Einhorn-Savit
in terms of the lattice formulation \cite{savit}.
It is noted  that if we view $\vec c$ as a gauge field coupled to the
vortex current, the gauge
symmetry is broken in Eq.(11), which corresponds to the non-conservation of
the vortex current, i.e.,
$\nabla \cdot
\vec j_V$ can be nonzero, due to the instantons.
After integrating over the field $\vec c$ in Eq.(11),
the partition function $Z$ is given by the integral over the vortex
configurations $\{ \vec j_V \}$
including both the vortex loop  and open vortex segments
the two ends of which are instanton and anti-instanton \cite{savit}.
\begin{equation}
Z = Z_0 \sum_{ \{ \vec j_V \} } \exp [ \sum - 4 \pi^2 \kappa
j_{V \mu}(j) D_{\mu \nu}(j-k, m^2) j_{V \nu}(k) ]
\end{equation}
where $D_{\mu \nu}(j-k, m^2)$ is the propagator of the field $\vec c$
with the mass $m^2 = \kappa g$. This propagator is given as
\begin{equation}
D_{\mu \nu}(j-k, m^2) =
\biggl[ \delta_{\mu \nu} - { { \Delta_\mu \Delta_\nu } \over { m^2} }
\biggr]_j  D(j-k, m^2)
\end{equation}
where $D(j-k, m^2)$ satisfies
\begin{equation}
( - \Delta_\mu^2 + m^2 ) D(j-k, m^2) = \delta_{ jk}.
\end{equation}
The propagator decays exponentially as $D(j-k, m^2) \sim e^{ - m | j-k|}$,
and the interaction between the vortex segments are short range.

 Let us start with  a qualitative estimate of the free energy of the vortex
loop/segment
regarding the partition function as that of a classical
statistical mechanics. We first repeat the argument of Einhorn and Savit.
\cite{savit}  Let $L$ be the length of the
vortex loop/segement. Then the free energy is the sum of the energy cost
and the entropy as
\cite{savit}
\begin{eqnarray}
F_{\rm loop}  &=& 4 \pi^2 \kappa D(0; m^2) L - \ln \tilde{\mu}^L
\nonumber \\
F_{\rm segment}  &=& 4 \pi^2 \kappa D(0; m^2) L - \ln (\tilde{\mu}^L
L^\gamma  )
                   + 2 S_{\rm inst}
\end{eqnarray}
where $S_{\rm inst} = 4 \pi^2 g^{-1} D(0; m^2)$ is the action for a
instanton, and
$\tilde{\mu}$ is a number of order unity which depends on lattice and dimension.
The main difference between closed loop and open segment  is that
the two instanton action is added and entropy is enhanced by the factor
$L^\gamma$ $(\gamma>0)$ in the case of open segment \cite{degennes}.
However the leading $L$-linear terms are the same for both of
closed loop and open segment, and Einhorn-Savit concluded that
the proliferation of the vortices, i.e., the appearance
of the infinite length loop/segment, occurs at
$4 \pi^2 \kappa D(0; m^2) = ln\tilde{\mu}$.
This is the XY-like phase transition viewed in the duality picture.
However the above consideration neglects the possibility that the
long vortex loop/segment are cut by the instantons into small pieces, as
shown in Fig. 2c.
Let us consider that the total length $L$ is cut into $n$ pieces of
open segments. The free energy in this case is
\begin{equation}
F(L,n) =  (4 \pi^2 \kappa D(0; m^2)- \tilde{\mu} )L
 -  \gamma n \ln ( L/n) + 2 S_{\rm inst}  n
\end{equation}
Minimizing this with respect to $n$, we obtain the length of the pieces
as $L/n = e^{1 + 2S_{\rm inst}/\gamma}$, which is finite as $L \to \infty$.
Then due to the finite density instantons, the infinite length loop or
segment does not appear. Instead the vacuum is full of finite size
open segments and closed loops of vortex. The total length of these
loops/segments are infinite as the sample size $L \to \infty$
when $4 \pi^2 \kappa D(0; m^2)- \mu <0$, but
they are all finite size.
Therefore we conclude that the phase diagram of (2+1)D Abelian Higgs model
is given by Fig. 1(b), i.e., all the interior constitutes a single phase.
Analogy of the phase diagram with the Ising model is useful here.
In the duality picture, i.e., regarding the vortex field as the order
parameter, $\kappa$ can be regarded as the temperature $T$, and the
instanton density $e^{- {\rm const}/g }$ is the magnetic field $H$.
Then it is natural that the ordered state at  $1/g = \infty$, i.e.,
$H=0$, is the isolated line while all the other phase diagram is
connected to the high temperature symmetric phase.
Actually this analogy becomes more clear when one consider the
path integral formulation of the the Ising model under magnetic field
$H$ \cite{degennes,cloi}.
The partition function is represented by the sum over the closed loop and
the open segment ended at the magnetic field vertex $H$.
This is exactly similar to the present case where the
vortex segment terminates at instantons except that the vortex
loop/segment  has a direction and instanton has the $\pm$ topological charge.
Now what is the nature of this single phase ?
One crutial question is if the Meissner effect for $\vec a$ remains or not.
It should be noted that the magnetic field $\vec b$ is tied to the
vortices. In the case of closed loop, however, the net magnetic field
$\vec b$ is zero. The open segment,
on the other hand, is a magnetic dipole which has net
$\vec b$. Therefore once the instanton fugacity is
nonzero and there are vortex open segments,
the magnetic field $\vec b$ can penetrate into the sample, i.e., the Meissner
effect disappears and the gauge field $\vec b$ becomes massless.
This corresponds to the ``dielectrics'' of the magnetic
charge and is illustrated in Fig.3b. The instantons and anti-instantons appear
to be bound
into pairs. However this state is not stable because once the Meissner
effect is gone, the confinement between the instanton and
anti-instanton also disappears, and the magnetic dipole is liberated into
free magnetic charges. This is the ``metal'' of the magnetic charge, and
again the magnetic field $\vec b$ can not penetrate into the sample due to
the screening. This is shown in Fig. 3c. This massive gauge field $\vec b$
corresponds to
the
confining phase of the pure gauge model discussed by Polyakov \cite{pol}.
Therefore the state in the interior of the phase diagram in Fig. 1(b)
continues smoothly not to the Higgs phase but to the pure gauge model, and
should be called ``confinement''.

In order to substantiate the above consideration, we now go to the second
quantization formulation of the vortex system.
Let us first devide the gauge field $\vec c$ into the transverse and
longitudinal parts, i.e.,
$\vec c = \vec c_{\perp} + \nabla \phi$. Then the action eq.(11) is written as
\begin{equation}
S = \int d^3 x \biggl[ { 1 \over 2 \kappa} (\nabla \times \vec c_\perp)^2 +
{ g \over 2} ({\vec c_\perp})^2 + { g \over 2} ({\nabla \phi})^2
+ i 2 \pi \vec c_\perp  \cdot \vec j_V +  i \phi \cdot \rho_M
\biggr]
\end{equation}
where we have performed a partial integration and identified $\nabla \cdot
\vec{j}_V$ with the instanton
density $\rho_M$.  We can view Eq. (18) as describing world-lines of
vortex-particles which are coupled
to the gauge field by the $i 2\pi \vec{c}_\perp \cdot \vec{j}_V$ term.  The
vortices may be created and
annihilated at instantons located at $x_1 \cdots x_n$ and anti-instantons
located at $y_1 \cdots y_m$.
Alternatively, we can write the action in terms of the second quantized
vortex field $\psi_V$.  The
action is given by
\begin{equation}
S_V = \int d^3 x \biggl[ { 1 \over 2 \kappa} (\nabla \times \vec c_\perp)^2 +
{ g \over 2} ({\vec c_\perp})^2 + { g \over 2} ({\nabla \phi})^2
+ \psi_V^\dagger { 1 \over 2}[ - K ( \nabla + i \vec c_\perp )^2
   + M^2] \psi_V  + u ( \psi_V^\dagger \psi_V)^2 \biggr]
\end{equation}
where $M^2 = 4 \pi^2 \kappa D(0; m^2)- ln\tilde{\mu}$, and $u$ represents the
short range repulsion between the vortex segments. The gradient term
comes from the extra cost of the action when the vortex line deviates from
the straight line.  This step is standard in the duality mapping.  The
novel feature of creation and
annihilation of vortices can be included by summing over all instanton
configurations as follows,
\begin{eqnarray}
Z &=& \int D \psi_V^\dagger D \psi_V D \vec c_\perp D \phi
 \sum_{ n, m = 0}^{\infty}
\int { { d x_1 \cdot \cdot  d x_n } \over {n !}}
\int { { d y_1 \cdot \cdot  d y_m } \over {m !}}
\nonumber \\
&\times&
( z \psi_V^\dagger (x_1) e^{ i \phi(x_1) } ) \cdot \cdot
( z \psi_V^\dagger (x_n) e^{ i \phi(x_n) } ) \times
( z \psi_V (y_1) e^{ -i \phi(y_1) } ) \cdot \cdot
( z \psi_V (y_m) e^{ -i \phi(y_m) } ) e^{- S_V}
\end{eqnarray}
where
$z$ is the fugacity of the instantons which is roughly given as
$z \sim e^{- S_{\rm inst}} $.
$n, m$ are the number of instantons and anti-instantons, but only the
term $n=m$ survives when one integrates over $\psi_V, \psi^\dagger_V$.
As in the usual Coulomb gas mapping, the summation over $n, m$
in eq.(20) can be
done, and our final result for the action when recovering the
original gauge field $\vec a$ is given as
\begin{eqnarray}
S &=& \int d^3 x \biggl[ { 1 \over 2 \kappa} (\nabla \times \vec c_\perp)^2
+ { 1 \over {2 g} } (\nabla \times \vec a)^2
- i \vec c \cdot \nabla \times \vec a
\nonumber \\
&+&  { 1 \over 2} \psi_V^\dagger [ - K ( \nabla + i \vec c_\perp )^2
   + M^2] \psi_V  + u ( \psi_V^\dagger \psi_V)^2
- z (\psi_V^\dagger e^{ i \phi} + \psi_V e^{ -i \phi} )
\biggr]
\end{eqnarray}
where it should be noted again that $\vec c = \vec c_\perp + \nabla \phi$
and $\vec a = \vec a_0 + \vec a_{\rm inst}$.
The gauge transformation $\psi_V \to \psi_V e^{ i \phi}$,
$\psi_V^\dagger  \to \psi_V^\dagger  e^{ - i \phi}$ eliminates the
exponential factor in the $z$-term, and also replaces  $\vec c_\perp$
by $\vec c$ in the minimal coupling term.
We have
\begin{mathletters}
\begin{eqnarray}
Z & = & \int D\psi_V^\dagger D\psi_V D\vec{c}D\vec{a}e^{-S} \\
S & = & \int d^3 x  \left\{
\frac{1}{2 \kappa} (\nabla \times \vec{c})^2 +
\frac{1}{2g} (\nabla \times \vec{a})^2 - i\vec{c} \cdot \nabla \times \vec{a}
\right.
\nonumber \\
&+& \frac{1}{2} \psi_V^\dagger \left[ -K(\nabla +
i\vec{c})^2 + M^2  \right] \psi_V + u(\psi_V^\dagger \psi_V)^2 \nonumber \\
& &                             \left. \rule{0in}{.25in}        -
z (\psi_V^\dagger + \psi_V) \right\}
\end{eqnarray}
\end{mathletters}
Except for the last term, Eq.(22b) is the standard duality representation
of the abelian
Higgs model. It is useful to recall that vortices in the field $\psi_V$
corresond to world
lines of the original bosons. Condensation of the vortex field $\psi_V$
means the absence of
Bose condensation and vice versa. The last term in Eq. (22b) represents the
effect of the
instantons and is
the main new result of this
paper.  We note that it takes the form of an external field coupled to the
vortex field $\psi_V$.  This
is analagous to the Josephson coupling to an external superfluid with an
order parameter $z$. The external
order parameter will induce a nonzero order parameter $<\psi_V > \neq 0$,
even when $M^2 > 0$.  [Note we
have fixed the gauge and the nonzero order parameter is in the particular
fixed gauge.]  In the first
quantized picture of vortex loops and segments, this can also be understood
as follows.  Consider the
correlation function $C(\vec{x},\vec{x}^\prime) = < \psi_V^\dagger
(\vec{x})\psi_V(\vec{x}^\prime) > $.
In the absence of instantons, the points $\vec{x}$ and $\vec{x}^\prime$ are
connected by a vortex line
and long-range order is possible only when an infinite vortex line has zero
energy, i.e., $M^2 < 0$.
However, with finite $z$, an instanton and an anti-instanton appear near
$\vec{x}^\prime$ and $\vec{x}$
and create two finite segments, so that $C(\vec{x},\vec{x}^\prime)$ reaches
a finite value even as the
separation between $\vec{x}$ and $\vec{x}^\prime$ goes to infinity.

Recall that in the duality picture, $< \psi_V > \neq 0$ means that the
original boson is not
bose-condensed.  Thus we expect that the effect of the $z(\psi_V +
\psi_V^\dagger)$ term is to destroy
the Meissner effect of the original Abelian-Higgs  theory.  We check this
by an explicit calculation of
the gauge field correlation function.
Let us represent $\psi_V$ as
$\psi_V = \psi_0 e^{ i \varphi}$. Then the
action for the vortex field becomes
\begin{equation}
S_{\rm vortex} = \int d^3 x  { 1 \over 2}[
K\psi_0^2 ( \nabla \varphi + \vec c )^2
         - 4 z \psi_0 \cos \varphi ]
\end{equation}
which is the sine-Gordon model in (2+1)D.
According to the analysis of the pure gauge model by
Polyakov \cite{pol},
the fugacity $z$ is always relevant and $\varphi$-field is
massive. Replacing the $\cos$-term by the effective quadratic term as
\begin{equation}
S_{\rm vortex} = \int d^3 x { 1 \over 2}
[ K\psi_0^2 ( \nabla \varphi + \vec c )^2
        +  2 z_{\rm eff} \psi_0 \varphi^2 ]
\end{equation}
which gives the mass $m_0 = \sqrt{ 2 z_{\rm eff}/ (K\psi_0)}$ of the
$\varphi$-field corresponding to the screening.
After integrating over $\varphi$, we obtain the effective action for the
gauge fields $\vec c$, $\vec a$ and the external electromagnetic field
$\vec A$ as
\begin{eqnarray}
S &=& { 1 \over 2} \sum_q \sum_{\mu \nu}
\biggl[
\biggl( { {q^2} \over \kappa}
\biggl( \delta_{\mu \nu} - { { q_\mu q_\nu} \over { q^2} } \biggr)
+ K\psi_0^2 \biggl( \delta_{\mu \nu} - { { q_\mu q_\nu} \over {q^2 + m_0^2} }
\biggr) \biggr) c_\mu(q) c_\nu(-q)
\nonumber \\
&+&  { 1 \over g}
\biggl( \delta_{\mu \nu} - { { q_\mu q_\nu} \over { q^2} } \biggr)
 a_\mu(q) a_\nu(-q)
- i \delta_{\mu \nu}
[( a_\mu(q)+ A_\mu(q) ) c_\nu(-q) +c_\mu(-q) ( a_\nu(q)+ A_\nu(q) ) ]
\biggr]
\end{eqnarray}
After integrating over $\vec c$, we obtain the propagator of the gauge
flux $\vec b = \nabla \times \vec a$ as
\begin{eqnarray}
< b_\mu(q) b_\nu(-q) >  &=&
{ 1 \over { g + K\psi_0^2 + q^2/\kappa} }  \delta_{\mu \nu}
\nonumber \\
&-& q_\mu q_\nu
{  { g [ m_0^2+\kappa K\psi_0^2 + q^2] } \over
{[m_0^2(K\psi_0^2+g)/g + q^2] [\kappa(K\psi_0^2+g)+q^2]} }
\end{eqnarray}
It should be noted that  while
the pole $1/q^2$ disappeared and the gauge field is massive, there is no
Meissner effect because Eq. (26)
is not proportional to $q^2$ as in Eq. (2). Rather, the small $q$ behavior
is essentialy the same as
that in the confining phase in the pure gauge model shown in
Eq. (7) \cite{pol}.
After integrating over the $\vec b$ field, the effective action for the
external e.m. field  is $ \sim (\nabla \times \vec A)^2$,
which means that the system is insulating.  This is perhaps not surprising
if viewed in the strong
coupling limit $g \rightarrow \infty$.  Then the gauge field does not have
its own dynamics and serves to
impose the constraint of integer occupation at each lattice site.  The
bosons are just frozen into place
on each site, resulting in an insulator.  This Mott insulator phase appears
to extend to include the
entire phase diagram, as shown in Fig. 1(b), with the exception of the line
$\frac{1}{g} = 0$.

 At finite temperature $T$, the imaginary time axis becomes finite, i.e.,
$[ 0, \beta=1/T]$, in eq.(23). Therefore eq.(23) desciribes the
sine-Gordon model in 2D in the long wavelength limit.
Therefore we expect the KT transition, i.e., confinement-deconfinement
transition, at some critical temperature $T_c$.

\section{Bosons with Non-integer Density}
The Abelian Higgs model in eq.(1) corresponds to the case of integer
boson filling for each site. The deviation from it is taken care of by
introducing the winding number of the phase $\theta$ along the time
direction \cite{dhl}. In the dual picture, the boson density is
represented by the
$z$-component of the gauge flux $\nabla \times \vec c$, and
the deviation from the integer filling is represented by adding the
term  $ - \mu (\nabla \times \vec c)_z $ to the action eq.(22). Here the
chemical potential $\mu$ acts as the magnetic field.
Therefore  the chemical potential and the vortex condensation
compete with each other as in the case of the superconductor in a magnetic
field.
The only new aspect here is the instanton term $z(\psi_V + \psi_V^\dagger)$.
When $\mu = 0$,
this term induces the vortex condensation $<\psi_V >$ even if $M^2$
is positive and large.
For $\mu \neq 0$, we may be tempted to consider an Abrikosov vortex state
(of the vortex field $\psi_V$)
in analogy with type II superconductors. An ordered array of such vortices
correspond to a
Wigner crystal of boson. [19]  However the phase of $\psi_V$
changes by $2\pi$ around each
vortex and we cannot gain the Josephson energy from the term $z(\psi_V +
\psi_V^\dagger)$. We conclude that the Wigner crystal is suppressed by the
instantons. A second idea
is to consider the analogy of the intermediae state in type I
superconductors, where the stable
configuration is the laminar structure \cite{deg2}.  
In type I superconductors the
surface energy between the
normal and superconducting regions is positive.  The surface energy is
proportional to the size along the
$z$-axis ($\beta$ in the present context) and the spacing of the laminar
structure is macroscopic in
size.  In the appendix, we perform a Ginzburg-Landau calculation of the
surface energy, and find that in
contrast to usual type I superonductors, it is negative in the case $M^2 > 0$,
i.e., when the superconductivity
is induced by Josephson coupling.  This implies that the straight interface
is unstable, and the laminar
phase will break up.  One possibility is that the system breaks up into
patches where $< \psi_V > \neq
0$, separated by regions where $< \psi_V > = 0$.  (This can be viewed as
the complement of the Abrikosov
vortex state.)  The order parameter can be real
in each patch, gaining an
extensive Josephson energy from the term $- z(\psi_V + \psi_V^\dagger)$.  The
magnetic field $\nabla \times
\vec{c}$ can penetrate the normal region and partially penetrate the
patches.  This state can maximize
the surface energy gain for a fixed patch area and the patches will form some
ordered structure.  In the
original boson representation, the absence of long-range order in $\psi_V$
means that the bosons form a
superfluid with Meissner effect.  The instantons become irrelevant in this
sense, in contrast to the $\mu
= 0$ case.  The effect of the instanton is to cause a periodic modulation
of the boson density
(corresponding to the modulation of the magnetic field $\nabla \times
\vec{c}$ by the patches).  This
modulation is weak for $\kappa$ large ($M^2$ positive  and large) and grows
with decreasing $\kappa$. An ordered array of patches lead to a kind of
incommensurate
order. For
very small $\kappa$, $M^2 < 0$ and there is a strong tendency for the order
parameter $\psi_V$ to form in
the dual picture.  In this case the interface energy may become positive
and we cannot rule out a laminar
picture.  In the original boson picture this corresponds to stripes of
superfluids separated by Mott
insulators.  The transition between the stripe phase depends on details of
the parameters and we have not
attempted to work it out quantitatively. However, since stripes occur only
for positive interface energy in our model, the stripe size in expected to 
be macroscpic , by ananogy
with the laminar phase of type I superconductors.
Microscopic stripes might occur when $M^2<0$ and other 
interactions such as long range Coulomb
forces and commensurability energy are introduced.
We have not examined this issue in this paper, and this
is left for future studies.

\section{Other Order Parameters}

Up to now, we consider the charge $e$ bosons coupled to the gauge field.
However other types of order parameter appear  in gauge models of
high-Tc superconductors, which is the subject of this section.

First we consider the case of $q=2e$, which corresponds to the spinon
pairing order parameter coupled to the U(1) gauge field.
Here we take the unit where $2e=1$ and the flux quantization
is reduces to half.
Then the instanton becomes the end point of two vortices, which
modifies the $z$-term in eq. (22b) as
\begin{eqnarray}
S & = & \int d^3 x  \left\{
\frac{1}{2 \kappa} (\nabla \times \vec{c})^2 +
\frac{1}{2g} (\nabla \times \vec{a})^2 - i\vec{c} \cdot \nabla \times \vec{a}
\right.
\nonumber \\
&+& \frac{1}{2} \psi_V^\dagger \left[ -K(\nabla +
i\vec{c})^2 + M^2  \right] \psi_V + u(\psi_V^\dagger \psi_V)^2 \nonumber \\
& &                             \left. \rule{0in}{.25in}        -
z(\psi_V^\dagger \psi_V^\dagger + \psi_V \psi_V ) \right\}.
\end{eqnarray}
It is noted that  $z$-term is the quadratic term
and does not necessarily enforce the condensation of $\psi_V$,
Therefore two possibilities arises in this case.

\begin{description}
\item [(I)] Single vortex condensation, i.e., $<\psi_V> \ne 0$.
In this case the quantized charge, i.e., the integral of
$( \nabla \times \vec c_\perp )_z$
is $2e$ and the single charge $e$ can not appear. Therefore
the charge $e$ is confined.

\item[(II)] Vortex pair condensation, i.e.,
$<\psi_V \psi_V > \ne 0$ while $<\psi_V> = 0$.
In this case the quantized charge is
reduced to half, i.e., $e$.
Therfore the confinement of the charge $e$ does not occur.
\end{description}

Fig. 4 shows the phase diagram for $q=2e$, where the above two
possibilities correspond to I and II, respectively.
As for the small fluctuation of $\vec a$ is concerned,
there occurs no Meissner effect in both phases, although $\vec a$
is massive due to the confinement.
Therefore both phases is better called confining phase.
What distinguished these two phases is the discrete $Z_2$ symmetry.
\cite{frad} It has been discussed that
the limit $\kappa = \infty$ is the Ising gauge model, which shows
confinement and deconfinement transition at some critical value of
$g = g_c$ \cite{kogut}.

Secondly we study the two species of bosons $b_1$, $b_2$ coupled to the
gauge field with opposite charges $e$ and $-e$, respectively.
This situation occurs in the staggered flux state of an SU(2) formulation
for underdoped region. [21]
In this case the dual model is given by
\begin{eqnarray}
S & = & \int d^3 x  \{
\frac{1}{2 \kappa} [ (\nabla \times \vec{c_1})^2 +(\nabla \times \vec{c_2})^2]+
\frac{1}{2g} (\nabla \times \vec{a})^2
\nonumber \\
&-& i\vec{c_1} \cdot \nabla \times (\vec{a} + \vec{A_1})
  - i\vec{c_2} \cdot \nabla \times (-\vec{a} + \vec{A_2})
  - i \mu  ( \nabla \times (\vec{c_1} + \vec{c_2}) )_z
\nonumber \\
&+& \frac{1}{2}
\psi_1^\dagger \left[ -K(\nabla +i\vec{c_1})^2 + M^2  \right] \psi_1 +
\psi_2^\dagger \left[ -K(\nabla +i\vec{c_2})^2 + M^2  \right] \psi_2
\nonumber \\
&-& z(\psi_1^\dagger \psi_2 + \psi_2^\dagger \psi_1 )
+ u[ (\psi_1^\dagger \psi_1)^2 + (\psi_2^\dagger \psi_2)^2 ]
+ 2 w (\psi_1^\dagger \psi_1) (\psi_2^\dagger \psi_2) \}
\end{eqnarray}
where $c_i$ ($i =1,2$) is the gauge field representing the boson current
of $b_i$, $\psi_i$ is the corresponding vortex field and
$\vec A_i$ is the test field coupled to it.
The real electromagnetic field corersponds to $\vec A_1 = \vec A_2 = \vec A$.
It is noted here that again the instanton term ($z$-term) is
quadratic in $\psi$'s  and cannot induce nonvanishing $\psi_1$ and $\psi_2$
when $z$ is not
large enough. The Bose condensation of $b_1$ and $b_2$ should
occur in this case.. When $g$ and $z$ is large,
which is relevant to the high-Tc problem, the amplitudes of both $\psi_1$ and
$\psi_2$ are induced and we write
$\psi_i = \psi_0 e^{i \varphi_i}$. Here the singluar vortex configuration of
$\varphi_i$ is allowed, which corresponds to the original boson.
Then the effective action for the phase field $\varphi_i$ is given by
\begin{equation}
S_{\rm eff.} = \int d^3 x  \biggl[
{ 1 \over 2}K\psi_0^2
[ ( \nabla \varphi_1 + \vec c_1)^2 + ( \nabla \varphi_2 + \vec c_2)^2 ]
     - 2z \psi_0^2 \cos( \varphi_1-\varphi_2) \biggr]
\end{equation}
Here we define the symmetric and antisymmetric parts as
\begin{eqnarray}
\varphi_1 &=& \varphi_s + { 1 \over 2} \varphi_a
\nonumber \\
\varphi_1 &=& \varphi_s - { 1 \over 2} \varphi_a
\end{eqnarray}
and $\vec c_s$, $\vec c_a$ in a similar way.
Then eq. (29) is wirtten as
\begin{equation}
S_{\rm eff.} = \int d^3 x  \biggl[
{ 1 \over 2} K\psi_0^2
[ 2 ( \nabla \varphi_s+ \vec c_s)^2 +
  { 1 \over 2} ( \nabla \varphi_a + \vec c_a)^2 ]
     - 2z \psi_0^2 \cos( \varphi_a) \biggr]
\end{equation}
where only the antisymmetric part is coupled to the instantons.
Therefore the actoin for the antisymmetric part is the same as
that  in eq. (23), i.e.,  $\varphi_a$ is fixed by the $z$-term,
and the vortex of $\varphi_a$-field is forbidden. There is no Meissner effect
for the  gauge field
$\vec a$, although it is massive as in the pure gauge model.
This corresponds to the binding or confinement of the two species of bosons
$b_1$ and $b_2$ because
$( \nabla \times \vec c_a )_z$ is the difference between the
boson densities of $b_1$ and $b_2$.
This means that the single Bose condesation is suppressed.
The boson pairing condensation, on the other hand, is not
disturbed by the $z$-term.
Therefore when the field $\varphi_s$ is disordered,
we have the boson pair condensation and finite superfulidity density $\rho_s$.
Then the effective action for the test fields $\vec A_1$, $\vec A_2$ is
given after integrting over $\vec c$-fields as
\begin{equation}
S_{\rm A} = \int d^3 x  \biggl[
\rho_s (\vec A_1 + \vec A_2)^2 +
\chi_a ( \nabla \times (\vec A_1 - \vec A_2 ) )^2  \biggr],
\end{equation}
where $\chi_a$ is a diamagnetic susceptibility of the antisymmetric part.
The the system show the Meissner effect only for the symmetric test field
$\vec A_1 + \vec A_2$.
Therefore the system shows the Meissner effect to the
external electromagnetic field $\vec A$.

\section{Conclusions}

In this paper, we studied the interplay between the confinement and the
condensation of the order parameters.
At integer-filling of the bosons with charge $e$, there is only one phase
in (2+1)D with the XY-transition isolated only on the line $g = 0$.
The nature of this so-called Higgs-confinement phase is the same as the
confining phase of  the pure gauge model, and no Meissner effect for the
gauge field occurs. This is because the instantons act as ordering field
for the vortex condensation. For non-integer filling, Bose condensation is
recovered for weak
coupling $g$.  However, the Bose condensation and confinement
compete with each other, and this competition leads to phase separation for
strong coupling.

In this paper we have focused our attention to the problem of bosons
coupled to gauge
fluctuations.  It
is only a first step towards addressing the problem which arises out of the
gauge theory
formulation of the high T$_c$ problem, which involves both fermions and
bosons coupled to
gauge fields.  Nevertheless, we would like to put the present work in the
context of the
high T$_c$ problem and attempt to draw a few inferences.  The effect of the
fermion is
two-fold.  First, if the gauge field is confining, it allows the
possibility of confining
fermion-antifermion pairs to form spin excitations and confining fermion
and boson to re-constitute the physical electron.  
The former is believed to happen in the half-filled
case where the AF ordering may be described as chiral symmetry breaking and
confinement of
Dirac fermions.\cite{marston,kim}  Secondly, the presence of massless Dirac
particles changes
the dynamics of the gauge field and, in general, it would not take the
Maxwellian form assumed
in this paper.  We can divide the doping region of the phase diagram into
three regimes:

\begin{itemize}
\item[(i)] {\it Doping into the AF} $(x \ll 1)$.  Here the starting point
is a $\pi$-flux
phase for the fermions with a Dirac spectrum.  The gauge propagator is
proportional to
$\sqrt{q^2}$ instead of $q^2$ in the Maxwell theory.\cite{ioffe}  The
instantons have
logarithmic interaction \cite{ioffe} and undergo a Kosterlitz-Thouless
transition in the $2 +
1$ dimension as a function of $N$, the number of fermion flavors.  It is
believed that the
physical case of $N=2$ lies on the disordered side of this transition, so
that the instanton
gas behaves as free gas (as opposed to instanton anti-instanton bound
pairs).  Since our
consideration is based on assuming the existence of the free instanton gas,
this would be the
case where our consideration has the best chance of being applicable.
Nevertheless, we still
have not included the possibility of bosons combining with fermions to form
physical holes in
an AF background.  This would correspond to the formation of small Fermi
liquid pockets in a
reduced Brillouin zone.  Leaving this possibility aside, we can conclude
from the results of
Section III that instantons suppress the formation of a Wigner crystal of
doped holes.
Furthermore, the possibility of phase separation into microscopic patches
is interesting, in
that it suggests incommensurate structures which appear experimentally in
this part of the
phase diagram.  However, the superfluid state that appear in our picture
does not appear to
resemble $d$-wave pairing, as long as the fermions remain confined in the
AF state.  It is
also interesting that phase separation into larger scale laminar domain is
a possibility.
Finally in the SU(2) formulation, the result of Section IV suggests the
possibility of
bosons forming a pairing state, leading to a
co-existence of
superconductivity and AF.
\end{itemize}

\begin{itemize}
\item[(ii)] {\it Underdoped region}.  Here the normal state is the
pseudogap state which is described as $d$-wave pairing of fermions or 
a staggered flux phase.\cite{wen}  Again
initially the fermion spectrum is Dirac and the gauge propagation is
proportional to $\sqrt{q^2}$ and it is not clear that the present paper is 
applicable. Nevertheless, we can
ask whether the low temperature phase is a confinement phase where
instantons are free and
play an important role.  There are three possible scenarios for the onset
of the low temperature superconducting phase.  
The first is a binding of fermions with 
bosons to form
physical quasiparticles.  Since the fermions are already paired, a
superconducting state
appears.  This possibility is clearly beyond the scope of the present work.
The second and
third possibilities are the Bose condensation of single bosons in the U(1)
formulation, or
the pairing of two kinds of bosons in the SU(2) formulation.\cite{wen}  The
latter problem is
treated in Section IV.  What we learn from the present study is that
instantons tend to
suppress Bose condensation when the coupling constant $g$ is larger.
Furthermore, instantons
favor the binding of the two species of SU(2) bosons to form pairs which
then condense,
leading to a $d$-wave superconductor ground state.
\end{itemize}

\begin{itemize}
\item[(iii)] {\it The overdoped region}.  Here the high temperature phase
is the strange
metal phase and it has been argued that it is a de-confining phase due to
dissipation in the
gauge field dynamics.\cite{nag2}  The low temperature Fermi liquid phase is
best described as
a confinement of fermions and bosons.  These are clearly outside of the
scope of the present
paper.
\end{itemize}

The authors acknowledge X.G. Wen, C. Mudry, E. Fradkin, R.B. Laughlin,
T. K. Ng, and P.W. Anderson for fruitful discussions.
N.N.  is supported by Priority Areas Grants
and Grant-in-Aid for COE research
from the Ministry of Eduction, Science, Culture and Sports of Japan,
P.A.L. acknowledges the support by NSF-MRSFC grant number DMR-98-08941.
\appendix
\section{Surface Tension}
In this appendix, we show the calculation of the surface tension between the
normal and superconducting regions of the dual superconductor
following ref. \cite{lif}.
The free energy of the dual superconductor is given by
\begin{equation}
F = \int d^2 r \biggl[ { 1 \over 2 \kappa'} (\nabla \times \vec c_\perp)^2 +
 \psi_V^\dagger { 1 \over 2}[ - K ( \nabla + i \vec c_\perp )^2
   + M^2] \psi_V  -z(\psi_V + \psi_V^\dagger) +
   u ( \psi_V^\dagger \psi_V)^2 \biggr]
\end{equation}
where $ \kappa'^{-1} = \kappa + (d n/d \mu)^{-1}$ ( $d n/d \mu$:
the charge compressibility ), and the classical ( time-independent )
configuration is assumed.
The Ginzburg-Landau equations are obtained by taking the variation
with respect to $\delta \psi^\dagger_V$ and $\delta \vec c$.
\begin{equation}
{ K \over 2} (- i \nabla - \vec c_\perp)^2 \psi_V + { M^2 \over 2} \psi_V
 + u | \psi_V |^2 \psi_V = z
\end{equation}
\begin{equation}
 \nabla \times (\nabla  \times \vec c_\perp) = \kappa' \vec j_V
\end{equation}
where
\begin{equation}
{\vec j_V}={K \over 2i} ({{{\psi_V}^\dagger} \nabla \psi_V}-
(\nabla {\psi_V}^\dagger) \cdot \psi_V)
- A{|\psi_V|}^2 {\vec c_\perp}
\end{equation}
Now we consider the case of instanton driven dual superconductivity.
Namely the $z$-term is the driving force of the vortex condensation
and $M^2> 0$ .
Then we assume $M^2$ is large enough and $u (\psi_V^\dagger \psi_V)^2$
term can be neglected.
In the absence of the magnetic field ${\vec c_\perp}$ ,
$ \psi_V =\psi_{V0} ={{2z} \over M^2}$ ,
and the free energy measured from that in the normal state $F_{n 0}$
is given by
\begin{equation}
F-F_{n0}=-V {2z^2 \over M^2} \equiv -V {{H_c}^2 \over 2\kappa'},
\end{equation}
where $H_c$ is the thermodynammic critical field
and $V$ is the volume of the system.
Assume that the interface between the superconducting and normal regions is
localized near $x=0$ , and $x > 0$ region is superconducting.
Physical quantities depend only on $x$, and we choose the Coulomb gauge
${\nabla \cdot {\vec c_\perp} =0}$.
Therefore $ \partial_x {c_\perp}_x=0$, and we put ${c_\perp}_x=0$ .
The boundary condition is
\begin{eqnarray}
(\nabla \times {\vec c_\perp})_z &=&{ {dc_{\perp y} } \over dx}=H_c,
\ \ \ \ \ \
\psi_V =0 \ \ \ \ \ \ \ \ \ x\rightarrow - \infty
\nonumber \\
(\nabla \times {\vec c_\perp})_z &=&{ {dc_{\perp y} } \over dx}= 0,
\ \ \ \ \ \ \
\psi_V =\psi_{V0} \ \ \ \ \ x\rightarrow +\infty
\end{eqnarray}
Here we introduce normalized quantities.
\begin{eqnarray}
x &=& x/\lambda,
\nonumber \\
\psi &=&\psi_V/\psi_{V0},
\nonumber \\
c&=&c_{\perp y}/(H_c\lambda),
\end{eqnarray}
where $\lambda $ is the penetration depth and is given by
$\lambda =( 4K \kappa' Z^2 /M^4)^{-1/2}$.
The correlation length $\xi $ is given by $\xi = \sqrt{K/|M^2|}$,
and we define the ratio ${\eta \equiv \lambda /\xi}$.
Using these normalized quantities, the GL equations become
\begin{equation}
\psi''=\eta ^2 [1-(c^2+1)\psi], \ \ \ \ \ \ \
c''=c \psi^2 ,
\end{equation}
where ${\psi''=d^2\psi /dx^2}$ etc.,
and the boundary condition is
\begin{eqnarray}
c'&=&1,\ \ \ \  \psi' =0 \ \ \ \  x\rightarrow - \infty
\nonumber \\
c'&=&0,\ \ \ \  \psi =1  \ \ \ \  x\rightarrow + \infty
\end{eqnarray}
It can be easily shown from eg.(A8) that
\begin{equation}
{1 \over \eta^2 } \psi'^2+(c^2+1)\psi^2-2\psi-c'^2=-1
\end{equation}
The surface tension $\alpha _{ns}$ is given as follows.
First define the free energy density ${\tilde f}$ under the magnetic field
$H$ as
\begin{equation}
{\tilde f} = f - {{H B} \over {\kappa'}}.
\end{equation}
Then  $\alpha _{ns}$ is given by
\begin{eqnarray}
\alpha _{ns} &=& \int_{ -\infty}^\infty dx
( {\tilde f} - {\tilde f_n} )
\nonumber \\
 &=& \int_{ -\infty}^\infty dx
\biggl[ { { (\nabla \times \vec {c_\perp})^2} \over {2\kappa'} }
+  {K
 \over 2} (|\psi'_V|^2 +{\vec c_\perp}^2 |\psi_V|^2)
+{M^2 \over 2}  |\psi_V|^2 - z(\psi_V +\psi_V^\dagger) -
{ {H_c (\nabla \times \vec c_\perp )_z} \over {\kappa'}}
+  { {2z^2} \over M^2} \biggr]
\nonumber \\
&=&{ {\lambda H_c^2} \over {2\kappa'}} \int_{-\infty}^\infty d x
[(c'-1)^2+ {1 \over \eta^2}  (\psi')^2 + (c'^2 +1)\psi^2-2\psi],
\end{eqnarray}
where ${\tilde f_n}$ is the ${\tilde f}$ in the normal state.
Using eg.(A10),
\begin{equation}
\alpha_{ns} ={ {\lambda H_c^2}  \over {\kappa'}}
\int_{-\infty}^\infty d x  c'(c'-1)
\end{equation}

Because the normalized magnetic flux density
$c'$ is $0 < c' <1$
in the interface region, the integral in eq. (A13) is negative and
$\alpha_{ns} < 0$.
Therefore we conclude that the surface tension is negative in the
instanton-driven dual superconductor.

\vfill
\eject
\noindent
Figure captions
\par
\noindent
\\
Fig. 1.
Phase diagrams for Abelian Lattice Higgs model with fundamental charge
in (3+1)D (a) and (2+1)D (b).
In (b) the phase transition is isolated along the XY-model line,
i.e., $g=0$. In the confinement phase, the gauge field is massive
due to instantons, althogh no Meissner effect occurs.
\par
\noindent
Fig. 2.
(a) Vortex loop, (b) a vortex segment, and (c) vortex segment broken into
pieces by the instantons. A
vortex segment terminates at the instanton and anti-instanton.
\par
\noindent
Fig. 3.
Three possible state for the magnetic charges.
(a) The vacuum state where only vortex loops of finite size exist.
    The gauge flux $b$ can not exist inside the sample, i.e., the Meissner
    effect occurs, and this is the superfluid state of the original bosons.
(b) The dielectrics with finite size vortex segments and loops.
    The magnetic field $b$ can penetrate into the  sample, and the
    Meissner effect vanishes. The string tension of the vortex then
    becomes zero, and the confinement of the instanton and anti-instanton
    disappears. Therefore this state is unstable to (c).
(c) The metallic state of the magentic charges.
    The metallic screening prevents the gauge field from penetrating
    into the sample. This is the confining state of the gauge field.
\par
\noindent
Fig. 4.
Phase diagram for $q=2e$. In region I the single vortex condensation
occurs and charge $e$ is confined. In region II only vortex pairs condense
and charge $e$ is not confined. In the limit $\kappa=\infty$, the
model is reduced to Ising gauge model, which shows a phase transition.
\\

\end{document}